\documentclass[journal,draftcls,onecolumn,12pt,twoside]{IEEEtranTCOM}
\normalsize
\usepackage{epsfig,cite}
\usepackage{amsmath,amssymb,setspace}
\usepackage{multirow,array}
\newtheorem{proposition}{Proposition}
\newtheorem{lemma}{Lemma}
\newtheorem{remark}{Remark}
\newtheorem{corollary}{Corollary}

\begin{document}
\title{Energy-Efficient Sensing and Communication of Parallel Gaussian Sources}
\author{Xi~Liu,~\IEEEmembership{Student Member,~IEEE,}
        Osvaldo~Simeone,~\IEEEmembership{Member,~IEEE,}
        and~Elza~Erkip,~\IEEEmembership{Fellow,~IEEE}
\thanks{X. Liu and E. Erkip are with the ECE Department, Polytechnic Institute of New York University, Brooklyn,
NY 11201. (Email: xliu02@students.poly.edu, elza@poly.edu).}
\thanks{O. Simeone is with the ECE Department, New Jersey Institute of Technology, Newark, NJ 07102. (Email: osvaldo.simeone@njit.edu).}
}
\vspace{-0.5cm}
\maketitle
\vspace{-2cm}
\begin{abstract}
\vspace{-0.5cm}
Energy efficiency is a key requirement in the design of wireless sensor networks. While most theoretical studies only account for the energy requirements of communication, the sensing process, which includes measurements and compression, can also consume comparable energy. In this paper, the problem of sensing and communicating parallel sources is studied by accounting for the cost of both communication and sensing. In the first formulation of the problem, the sensor has a separate energy budget for sensing and a rate budget for communication, while, in the second, it has a single energy budget for both tasks. Furthermore, in the second problem, each source has its own associated channel. Assuming that sources with larger variances have lower sensing costs, the optimal allocation of sensing energy and rate that minimizes the overall distortion is derived for the first problem. Moreover, structural results on the solution of the second problem are derived under the assumption that the sources with larger variances are transmitted on channels with lower noise. Closed-form solutions are also obtained for the case where the energy budget is sufficiently large. For an arbitrary order on the variances and costs, the optimal solution to the first problem is also obtained numerically and compared with several suboptimal strategies.
\end{abstract}
\vspace{-0.5cm}
\begin{IEEEkeywords}
\vspace{-0.5cm}
Wireless sensor networks, energy-efficient communication, quantization, rate-distortion theory.
\end{IEEEkeywords}
\vspace{-0.5cm}
\section{Introduction}
\vspace{-0.25cm}
Sensor networks consisting of battery-limited nodes need to be operated in an energy-efficient manner in order to attain a satisfactory lifetime. The energy consumption of a sensor device is due mostly to the tasks of sensing and communication. The sensing component consumes energy in the process of digitizing the information sources of interest through a cascade of acquisition, sampling, quantization and compression, while the communication component spends power to operate the transmit circuitry and the power amplifier. It is known that the overall energy spent for compression is generally comparable to that used for communication and that a joint design of compression and transmission is critical to improve the energy efficiency \cite{processingcost:Barr_Asanovic06}\cite{processingcost:Sadler_Martonosi06}. We refer to the energy cost associated with measurements and compression of sources as ``sensing cost''.
\subsection{Contributions}
\begin{figure}
\centering
\includegraphics[width = 2.5in]{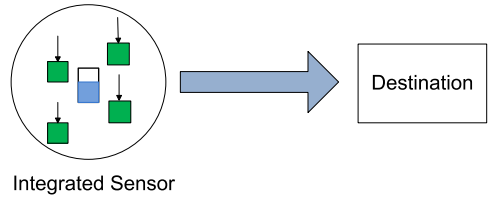}
\caption{Illustration of an integrated sensor device with multiple sensor interfaces sharing the same resource budget.}
\label{fig:integrated}
\end{figure}

In this paper, we consider an integrated sensor device consisting of multiple sensor interfaces \cite{processingcost:Wouters_Cooman_Puers94} that can simultaneously measure multiple information sources, as illustrated in Fig. \ref{fig:integrated}. Being part of the same device, the sensor interfaces share the same overall resource budget. Moreover, since the sensor interfaces have distinct hardwares and sensitivities, we assume that the sensing costs of different sources are generally different. Finally, for tractability, we model the sensing cost of a given source as being constant per source sample.

With sensing costs present, we aim at optimizing the resource (energy or rate) allocation for the integrated sensor of Fig. \ref{fig:integrated} so as to minimize the overall mean squared error distortion of the reconstruction of all the sources at the destination. We consider two types of resource constraints. In the first, the sensor has a given energy budget used for sensing and a separate rate constraint for communication (\textit{separate sensing/communication}). In the second, the sensor has an overall energy budget which is to be spent for both sensing and communication (\textit{joint sensing/communication}). Moreover, in the joint sensing/communication scenario, the sensed sources are assumed to be transmitted over orthogonal additive white Gaussian channels with different noise variances. This set-up can model a scenario in which different sensor interfaces of the integrated device are used at different times and, to avoid delay and buffer overflow, the measurements are transmitted over a time-varying channel to the destination as they are measured.

For the separate sensing/communication problem, we obtain a closed-form solution for the case where the sources with larger variances have lower sensing costs. This corresponds to a situation when sources with lower variances might require more energy-consuming sensor interfaces with higher sensitivity for sensing. When the source variances and the sensing costs are arbitrarily ordered, the optimal solution is obtained using numerical methods and is compared with several suboptimal strategies. For the joint sensing/communication problem, assuming that sources with larger variances not only have lower sensing costs, but also are transmitted over channels with lower noise variances, we obtain structural results on the optimal solution. Moreover, a closed-form solution is obtained for sufficiently large energy budgets.
\subsection{Related Work}
The joint design of compression and transmission parameters for energy efficiency has been investigated through the proposal of various algorithms for static scenarios in \cite{processingcost:Luna_Eisenberg_Berry_Pappas_Katsaggelos} \cite{processingcost:Akyol_vanderSchaar08} and for  dynamic scenarios in \cite{processingcost:Neely08} \cite{processingcost:Neely_Sharma08}. In particular, references \cite{processingcost:Neely08} and \cite{processingcost:Neely_Sharma08} proposed on-line algorithms that are able to choose among a finite number of compression options with different energy costs. Using Lyapunov optimization techniques, such algorithms can perform arbitrarily close to the minimal power expenditure for a given average distortion with an explicit trade-off in average delay. In \cite{processingcost:He_Wu06}, for wireless video sensors, an analytical model that characterizes the relationship between power consumption of a video encoder and the rate-distortion performance was developed. More recently, the problem of energy allocation over sensing and communication has been investigated for energy-harvesting sensors in \cite{processingcost:Castiglione_Simeone_Erkip_Zemen11}.
Finally, the model for the per-sample sensing cost in this paper is analogous to the per channel use processing cost used in \cite{processingcost:Massaad_Medard_Zheng04} to account for the transmitter processing power consumed by a wireless device. We remark that, in \cite{processingcost:Massaad_Medard_Zheng04}, when the processing energy cost is not negligible, it is no longer optimal to transmit continuously, but,  instead, bursty transmission becomes advantageous in terms of the achievable rate.

The rest of the paper is organized as follows. In Section II, we formulate the problems of interest. Then, Section III first derives the analytical optimal solution to the separate sensing/communication problem when the source variance and the sensing cost are ordered, and then addresses the same problem in the case of arbitrary parameters. In Section IV, the structure of the optimal solution to the joint sensing/communication problem is analyzed for the ordered case. Finally, we make some concluding remarks in Section V.
\section{Problem Formulation}
We consider a system in which a sensor measures $Q$ independent parallel Gaussian sources and communicates them to a single destination as shown in Fig. \ref{fig:integrated}. The $i$th source consists of $n$ independent and identically distributed (i.i.d.) samples with variance $\sigma_i^2$, $i=1,...,Q$. We assume that measuring each sample of the $i$th source entails a given sensing cost $\epsilon_{S,i}$ joules per source sample, which takes into account the energy spent for acquisition, sampling, quantization and compression. Note that, more generally, the energy costs associated with quantization and compression may depend on the compression rate and the target distortion level, as discussed in \cite{processingcost:Castiglione_Simeone_Erkip_Zemen11}. We do not pursue this more general model here for simplicity. We are interested in minimizing the overall average distortion $D$ of the reproduction of the sources at the destination. We consider two related problems. In the first (separate sensing/communication), we assume that the sensor has two resource budgets, an energy budget for sensing and a rate budget for communication. In the second (joint sensing/communication), instead, we consider the problem of allocating energy between the tasks of sensing and communications. Note that the second problem is in fact dual to the problem of minimizing the total energy consumed by the sensor subject to a given constraint on the allowed distortion level.
\vspace{-0.1cm}
\subsection{Separate Sensing/Communication of Parallel Sources}
\vspace{-0.2cm}
For the separate sensing/communication problem, we assume the sensor has an energy budget $E$ to be used exclusively for sensing of the $Q$ sources, and a total rate $R$ that can be allocated for communication. Both $E$ and $R$ are normalized by $n$ so that $E$ is the energy budget per source sample and similarly for $R$. When $E$ and $R$ are limited, it might not be optimal, or possible, to sense all the samples from all the sources. We assume instead that the sensor measures a fraction $\theta_{S,i}$, with $0\leq \theta_{S,i}\leq 1$, of samples from the $i$th source, and then sends a compressed version of them with rate $R_i$ ($R_i\geq 0$). Given the above, the mean square error (MSE) of the reconstruction for the $i$th source can be obtained as $D_i = \sigma_i^2 f(\theta_{S,i},R_i)$, where $ f(\theta_{S,i},R_i) = (1-\theta_{S,i})+\theta_{S,i}2^{-2R_i/\theta_{S,i}}$ if $\theta_{S,i}>0$, and $f(\theta_{S,i},R_i) =1$ if $\theta_{S,i}=0$ \cite{processingcost:Cover91}.

We define the sampling fraction vector and rate allocation vector as $\boldsymbol\theta_S = [\theta_{S,1}\; ...\; \theta_{S,Q}]^T$ and $\mathbf{R} = [R_1\;...\;R_Q]^T$, respectively. The problem of minimizing the total MSE is given by
\vspace{-0.2cm}
\begin{equation}
\min_{\boldsymbol\theta_S,\mathbf{R}}\; D(\boldsymbol\theta_S,\mathbf{R}) = \sum_{i=1}^Q \sigma_i^2 f(\theta_{S,i},R_i), \label{eqn:optprob}
\end{equation}
\vspace{-0.2cm}
subject to the sensing energy constraint $\sum_{i=1}^Q \theta_{S,i}\epsilon_{S,i}\leq E$ and the rate constraint $\sum_{i=1}^Q R_i \leq R$.
\vspace{-0.35cm}
\subsection{Joint Sensing/Communication of Parallel Sources}
\vspace{-0.25cm}
For the joint sensing/communication problem, the communication link is modeled as a collection of $Q$ orthogonal channels. We assume that the compressed version of the sensed samples from the $i$th source ($1\leq i\leq Q$) are transmitted over the $i$th channel, which is an independent complex Gaussian noise channel with noise variance $N_i$. Each channel consists of $n\tau$ channel uses, where $\tau$ is the channel-source bandwidth ratio for each source-channel pair. It is also assumed that the sensor has a joint energy constraint $B$ on the sensing and communication components. Similar to $E$ and $R$ in Section II-A, the energy $B$ is normalized by $n$ as well. The sensor measures a fraction $\theta_{S,i}$ of the samples of the $i$th source, and transmits the corresponding compressed samples with power $P_i$ over the $i$th channel. Since the compression rate for each sensed sample of the $i$th source is given by $(\tau/\theta_{S,i})\log_2(1+P_i/N_i)$, the MSE of the reproduction of the $i$th source at the destination can be obtained as $D_i = \sigma_i^2 h(\theta_{S,i},P_i)$, where $h(\theta_{S,i},P_i) = (1-\theta_{S,i})+\theta_{S,i} (1+P_i/N_i)^{-2\tau/\theta_{S,i}}$ if $\theta_{S,i}>0$, and $h(\theta_{S,i},P_i) = 1$ if $\theta_{S,i} = 0$.

We define the power allocation vector as $\mathbf{P} = [P_1\;...\;P_Q]^T$. The problem of minimizing the overall MSE is then given by
\vspace{-0.2cm}
\begin{equation}
\min_{\boldsymbol\theta_S,\mathbf{P}}\; D(\boldsymbol\theta_S,\mathbf{P}) = \sum_{i=1}^Q \sigma_i^2 h(\theta_{S,i},P_i) \label{eqn:optprob2}
\end{equation}\vspace{-0.2cm}
subject to the overall energy budget constraint $\sum_{i=1}^Q \theta_{S,i}\epsilon_{S,i}+\tau P_i\leq B$.
{\remark For the joint sensing/communication problem, if we have separate energy constraints on sensing of the $Q$ sources and on communication over the $Q$ channels, and if we allow the compressed version of the sensed samples from all sources to be transmitted across all parallel channels, then the problem reduces to the separate sensing/communication problem with rate $R$ given by the capacity of the channel made of the $Q$ parallel AWGN links subject to the transmit power constraint. Note that this capacity is obtained by water-filling \cite{processingcost:Cover91}.}
{\remark The above formulations can be extended to a more general case in which the destination wishes to minimize the weighted MSE distortion, i.e., the objective function is $\sum_{i=1}^Q w_i\sigma_i^2 f(\theta_{S,i},R_i)$ in (\ref{eqn:optprob}) or $\sum_{i=1}^Q w_i \sigma_i^2 h(\theta_{S,i},P_i)$ in (\ref{eqn:optprob2}), where $w_i>0$ is the weight for source $i$, with $i = 1,2$. The weight can in general account for the source priority. In this case, it is easy to see that it is enough to modify the variance of source $i$ as $w_i\sigma_i^2$ in order to convert the weighted MSE criterion to the standard MSE criterion considered throughout the paper. Hence, all solutions developed henceforth apply to weighted MSE distortion as well.}
\vspace{-0.35cm}
\section{Separate sensing and communication}
\vspace{-0.25cm}
This section considers the separate sensing/communication problem described in Section II-A. To facilitate the analysis, we divide the $Q$ Gaussian sources into $K$ classes with class $k$ ($1\leq k\leq K$) containing $q_k$ sources with the same variance $\sigma_k^2$. Without loss of generality, the variances are in descending order, i.e., $\sigma_1^2 > \sigma_2^2>...>\sigma_K^2$. Since each class can contain an arbitrary number $q_k$ of sources, we have strict inequalities among the variances. It is also assumed that sources in class $k$ have the same sensing cost $\epsilon_{S,k}$. In the following, we first analyze the optimal solution for the case when the sensing costs of $K$ classes are also ordered and then discuss the more general case.
\vspace{-0.25cm}
\subsection{The ordered variance/cost case}
\vspace{-0.25cm}
In this subsection, we assume that sources with larger variances have lower sensing costs, i.e., $\epsilon_{S,1} \leq ... \leq\epsilon_{S,K}$. Such an order would be valid if more energy-consuming sensor interfaces with higher sensitivities are required to measure sources with lower variances. Note that, while for the general case, the problem in (\ref{eqn:optprob}) can be shown to be convex, there is no closed-form solution, as will be discussed later. Focusing on the ordered case as described above allows us to obtain an analytical expression for the optimal solution and gain insights into the problem. By the convexity of function $D(\boldsymbol\theta_S,\mathbf{R})$, it is easy to see that we can assume the same fraction $\theta_{S,k}$ and rate $R_k$ are assigned to each source in the $k$th class\footnote{In fact, fixing all other parameters, function $D(\boldsymbol\theta_S,\mathbf{P})$ is Schur-convex with respect to the fractions of samples and the rates assigned to the sources in a class. Therefore, an equal fraction and rate allocation is optimal (see, e.g., \cite{processingcost:Marshall_Olkin79}).}.

For convenience, we divide the range of the energy $E$ into a sequence of intervals $\mathcal{E}_1 =(e_0,e_1]$, $\mathcal{E}_2 = (e_1,e_2]$,..., $\mathcal{E}_K = (e_{K-1},e_K)$, where $e_0 = 0$, $e_K = +\infty$ and $e_m = \sum_{i=1}^m q_i\epsilon_{S,i}$ for $1\leq m\leq K-1$, and divide the range of rate $R$ into a sequence of intervals $\mathcal{R}_1 = (r_0,r_1]$, $\mathcal{R}_2 = (r_1,r_2]$,..., $\mathcal{R}_K = (r_{K-1},r_K)$, where $r_0 = 0$, $r_K = +\infty$ and
\begin{equation}
 r_l = \frac{1}{2}\sum_{j=1}^l q_j \log_2\left(\frac{\sigma_j^2}{\sigma_{l+1}^2}\right), \quad 1\leq l\leq K-1. \label{eqn:rn}
\end{equation}
{\proposition For $K\geq 2$, assuming $\sigma_1^2 > ...> \sigma_K^2$ and $\epsilon_{S,1} \leq ... \leq\epsilon_{S,K}$, the optimal solution for the separate sensing and communication problem in Section II-A is obtained as follows. Given $E\in \mathcal{E}_m$ for some $1\leq m \leq K$,
\begin{enumerate}
\item If $R\in \mathcal{R}_l$ for some integer $l$ with $1\leq l\leq m-1$, then it is optimal to fully sample the first $l$ classes of sources, i.e., $\theta_{S,k}^* = 1$ for $1\leq k\leq l$,
    and to allocate rates as
     \begin{equation}
     R_k^* = \frac{1}{\sum_{j=1}^l q_j}\left(R+\frac{1}{2}\sum_{j = 1, j\neq k}^{l} q_j\log_2\left(\frac{\sigma_k^2}{\sigma_j^2}\right)\right),\label{eqn:R_k4}
    \end{equation}
     where $1\leq k\leq l$. Moreover, there is no need to sense the remaining $K-l$ classes of sources, i.e., $\theta_{S,k}^* = 0$ and $R_k^* =0$, for $l+1\leq k\leq K$.
\item If instead $R>r_{m-1}$ (or $R\in \bigcup_{l\geq m} \mathcal{R}_l$), then it is optimal to sample the first $m-1$ classes of sources fully, i.e., $\theta_{S,k}^* = 1$ for $1\leq k\leq m-1$, and the $m$th class for a fraction $\theta_{S,m}^* = \min((E-e_{m-1})/(q_m\epsilon_{S,m}),1)$, and to allocate rates as
\begin{equation}
 R_k^* = \frac{\theta_{S,k}^*}{\sum_{j=1}^{m} q_j\theta_{S,j}^*}\left(R+\frac{1}{2}\sum_{j = 1, j\neq k}^{m} q_j\theta_{S,j}^* \log_2\left(\frac{\sigma_k^2}{\sigma_j^2}\right)\right),\label{eqn:R_k3}
 \end{equation}
 where $1\leq k\leq m$. Moreover, there is no need to sense the remaining $K-m$ classes of sources, i.e., $\theta_{S,k}^* = 0$ and $R_k^* = 0$ for $m+1\leq k\leq K$.
\end{enumerate}
\label{thm:k3}
}
\vspace{-0.25cm}
\begin{proof}
The proof is based on solving the KKT conditions but special care must be taken since the objective function in (\ref{eqn:optprob}) is not continuously differentiable in the entire feasible set. Details of the proof are provided in Appendix A.
\end{proof}
In the zero sensing cost case, i.e., with $\epsilon_{S,1} = ... = \epsilon_{S,K} = 0$, we have $E\in \mathcal{E}_K = (0,+\infty)$, i.e., $m = K$ in Proposition 1. Hence, we have the following corollary.
{\corollary \label{cor:cor1} If $\epsilon_{S,1} = ... = \epsilon_{S,K} = 0$, the optimal solution is as follows: If $R\in \mathcal{R}_l$ for some integer $l$ with $1\leq l\leq K$, then it is optimal to fully sample the first $l$ classes of sources, i.e., $\theta_{S,k}^* = 1$ for $1\leq k\leq l$, and to set $\theta_{S,k}^* = 0$ for $l+1\leq k\leq K$, with rates $R_k^*$ as in (\ref{eqn:R_k4}) for $1\leq k\leq l$ and $R_k^* =0$ for $l+1\leq k\leq K$.}

{\remark If energy $E$ and rate $R$ are such that conditions in Case 1 of Proposition \ref{thm:k3} are satisfied, the optimal solution is not unique. This is because an amount $E-e_l$ of energy is left after fully sampling of the first $l$ classes that can be used to sense the remaining $K-l$ classes. In fact, the solution will remain to be optimal, if, instead of setting $\theta_{S,l+1}^*,...,\theta_{S,K}^*$ all to be zero, we set them to any values such that $0\leq \theta_{S,k}^* \leq 1$, $l+1\leq k\leq K$ and $\sum_{k=l+1}^K q_k\theta_{S,k}^*\epsilon_{S,k}\leq E-e_l$. The same discussion applies to Corollary \ref{cor:cor1}.}

Before we discuss the solution given in Proposition 1, we revisit the classical reverse water-filling approach, which solves the separate sensing and communication problem in (\ref{eqn:optprob}) in the case of zero sensing costs as in Corollary \ref{cor:cor1}. The interpretation below will also be useful in understanding the solution to the joint sensing and communication problem in Section V. With zero sensing costs, as stated in Corollary \ref{cor:cor1}, the solution only depends on the rate constraint $R$. Moreover, the solution to (\ref{eqn:optprob}) is obtained by solving the dual problem \cite{references:Bertsekas} \cite{processingcost:Boyd_Vandenberghe}
\vspace{-0.2cm}
\begin{equation}
\min_{R_k\geq 0} \left(\sigma_k^2 2^{-2R_k} + \alpha R_k\right),\label{eqn:int_rwf}\vspace{-0.2cm}
\end{equation}for each class $k$ where $\alpha$ is the Lagrangian multiplier (or ``rate price'') to be selected such that the total rate constraint $\sum_{k=1}^K q_k R_k = R$ is satisfied. Note that we can write
\vspace{-0.2cm}
\begin{equation}
\sigma_k^2 2^{-2R_k} = \int_0^{R_k}(-2\ln2)\sigma_k^2 2^{-2r}dr + \sigma_k^2,\vspace{-0.2cm}
\end{equation}so that the problem in (\ref{eqn:int_rwf}) can be recast as
\vspace{-0.2cm}
\begin{equation}
\min_{R_k\geq 0} \int_0^{R_k}(\alpha - (2\ln2)\sigma_k^2 2^{-2r}) dr =\min_{R_k\geq 0} \int_0^{R_k}w_k(r)dr, \label{eqn:recast1}
\end{equation}\vspace{-0.2cm}where we have defined $w_k(r)$ as $w_k(r) = \alpha - (2\ln2)\sigma_k^2 2^{-2r}$ and neglected the constant term $\sigma_k^2$. The product $w_k(r)dr$ can thus be interpreted as the marginal cost (rate price minus reduction in distortion) of adding an additional rate $dr$ when the currently assigned rate is $r$. For a given rate price $\alpha$, the solution of problem (\ref{eqn:recast1}) (and hence (\ref{eqn:int_rwf})) for any class $k$ with $w_k(0)<0$ is to increase the rate progressively until $w_k(r)$ becomes zero. The corresponding optimal rate is
\vspace{-0.2cm}
\begin{equation}
R_k^* = \left(\frac{1}{2}\log_2\left(\frac{(2\ln2)\sigma_k^2}{\alpha}\right)\right)^+,\label{eqn:Rk}\vspace{-0.2cm}
\end{equation}
where $(\cdot)^+$ denotes $\max(\cdot,0)$. Note that if the source variance is sufficiently small so that $w_k(0)\geq 0$, then no rate is assigned to the source at all. To obtain the optimal Lagrange multiplier $\alpha$ in (\ref{eqn:Rk}), we invoke the rate constraint $\sum_{k=1}^K q_kR_k = R$. It can be easily seen that, the optimal rate $R_k^*$ in (\ref{eqn:Rk}) is positive only for the $l$ classes of sources with the largest variances, where $l$ is such that $R\in \mathcal{R}_l$ and, moreover, $R_k^*$ can also be expressed as in (\ref{eqn:R_k4}) (see Appendix A).

Proposition \ref{thm:k3} states that, when the sensing costs are taken into account, the optimal solution in the ordered case entails sensing sources with the highest variances and then optimally allocating rates among the sensed sources using either the reverse water-filling procedure or a variation of it. Specifically, in case 1 of Proposition 1, that is, if $E\in \mathcal{E}_m$ with $1\leq m\leq K$ and $R\in \mathcal{R}_l$ with $1\leq l\leq m-1$, the first $l$ classes of sources are fully sensed and compression rates are assigned according to the classic reverse water-filling solution. Note that in this case, even though there is enough energy to sample more than $l$ sources, given the rate constraint, the optimal rate allocation only assigns positive rate to the first $l$ classes. Instead, in case 2 of Proposition 1, i.e., if $E\in \mathcal{E}_m$ and $R>r_{m-1}$, it is optimal to fully sample the first $m-1$ classes of sources, while the sources in the $m$th class are sampled only partially using the remaining energy. For the $m$th class, the optimal sampling fraction is equal to $\theta_{S,m}^* = \min((E-e_{m-1})/(q_m\epsilon_{S,m}),1)$, and the optimal rate is obtained, for a fixed rate price $\alpha$, by solving the dual problem
\vspace{-0.2cm}
\begin{equation}
\min_{R_m\geq 0} \left(\theta_{S,m}^*\sigma_m^2 2^{-\frac{2R_m}{\theta_{S,m}^*}} + \alpha R_m\right).\vspace{-0.3cm}
\end{equation}
Therefore, the marginal cost becomes $w_m(r)dr$ with $w_m(r) = \alpha - (2\ln2)\sigma_m^2 2^{-\frac{2r}{\theta_{S,m}^*}}$ and the optimal rate allocation $R_m^*$ is given by
\vspace{-0.2cm}
\begin{equation}
R_m^* = \left(\frac{\theta_{S,m}^*}{2}\log_2\left(\frac{(2\ln2)\sigma_m^2}{\alpha}\right)\right)^+.\label{eqn:Rm}\vspace{-0.3cm}
\end{equation}
Comparing with (\ref{eqn:Rk}), it is seen that rate assigned to each source in class $m$ is scaled by the fraction $\theta_{S,m}^*$. Moreover, from (\ref{eqn:optprob}), the distortion attained for each source in class $m$ is given by $\theta_{S,m}^* \hat{D}_m^*+(1-\theta_{S,m}^*)\sigma_m^2$, where $\hat{D}_m^*=\sigma_m^2 2^{-2R_m^*/\theta_{S,m}^*}$ is the normalized distortion for the sampled fraction of the source, while $\sigma_m^2(1-\theta_{S,m}^*)$ corresponds to the total distortion of the non-sampled fraction. By imposing the rate constraint $\sum_{k=1}^K q_kR_k=R$, as done above for the conventional reverse water-filling solution, we obtain (\ref{eqn:R_k3}) (see Appendix A).

\begin{figure}
\centering
\includegraphics[width = 4in]{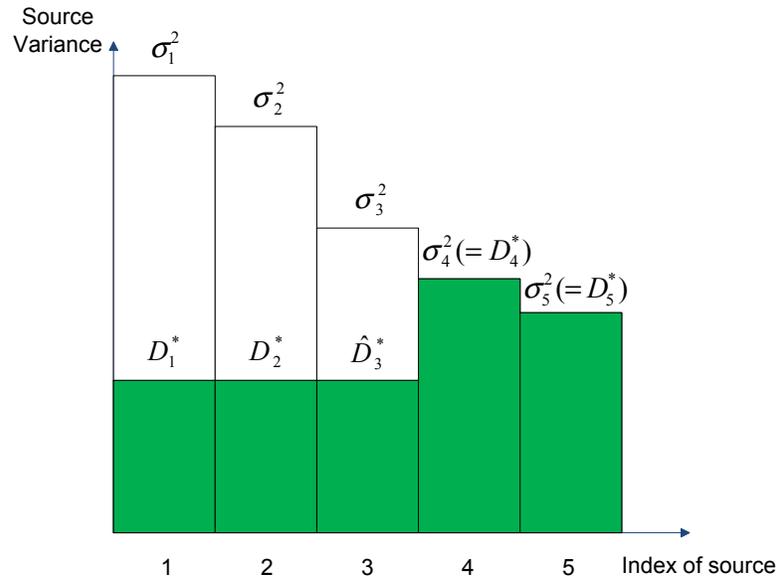}
\caption{Illustration of the optimal solution for case 2 of Proposition 1, where $K = 5$, $q_k = 1$ for $k = 1,...,5$ and $E$ and $R$ are chosen to satisfy $e_2<E< e_3$ and $R>r_4$.}
\label{fig:rwf}
\end{figure}

\vspace{-0.25cm}We pictorially illustrate the solution for case 2 of Proposition 1 in Fig. \ref{fig:rwf}, where we assume $K = 5$ and $q_k = 1$, $k=1,...,5$. In this example, the energy $E$ and the rate $R$ are assumed to satisfy $e_2<E<e_3$ and $R>r_4$. Thus, it is optimal to have source 1 and source 2 both fully sampled and have source 3 only partially sampled for a fraction $\theta_{S,3}^* = (E-e_2)/\epsilon_{S,3}$. The first two sources and the sampled fraction of source 3 are all described with the same distortion, i.e., $D_1^* = D_2^* = \hat{D}_3^* = \alpha/(2\ln2)$, where we recall that $\hat{D}_3^*$ is the average distortion only for the sampled fraction of source 3. The rate price $\alpha$ is set such that the sum constraint $R_1^* + R_2^* + R_3^* = R$ is satisfied. Since source 4 and source 5 are not sampled at all, they are assigned zero rates and thus the corresponding distortions are equal to their variances. Recall that in the zero sensing cost case, all the five sources are fully sampled and since $R>r_4$, all of them are described with the same per-source distortion, i.e., $D_1^* = D_2^* = ...= D_5^*$. Moreover, such per-source distortion (and thus the optimal rate price $\alpha$) would be larger than in the case of nonzero sensing costs shown in Fig. \ref{fig:rwf}, although the overall distortion in (\ref{eqn:optprob}) would be smaller.
\vspace{-0.25cm}
\subsection{The General Case}
\vspace{-0.25cm}
This subsection discusses the solution to the separate sensing/communication problem when the sensing costs $\epsilon_{S,k}$ are arbitrary. In this case, while the problem in (\ref{eqn:optprob}) is still convex, it appears prohibitive to obtain an analytical solution. Therefore, we resort to numerical methods. For non-differentiable objective functions, such as (\ref{eqn:optprob}), common convex optimization methods \cite{processingcost:Boyd_Vandenberghe}, like gradient descent and Newton's strategies, either do not apply or fail to converge. To avoid the non-differentiability of $D(\boldsymbol\theta_S,\mathbf{R})$ at points with $\theta_{S,k} = 0$, we assume that a fraction at least $\delta>0$ for each source is sampled, where $\delta$ is a small positive real number. The optimization problem remains the same as in (\ref{eqn:optprob}) except that constraint $0\leq \theta_{S,k}\leq 1$ is replaced with $\delta \leq \theta_{S,k}\leq 1$ for $1\leq k\leq K$. With such a modification, function $D(\boldsymbol\theta_S,\mathbf{R})$ becomes not only convex but also continuously differentiable over the new constraint set. The optimal solution to the modified problem, which can approximate that of the original problem in (\ref{eqn:optprob}) well when $\delta$ is made small, can be obtained numerically using common convex optimization methods. 

\begin{figure}
\centering
\includegraphics[width = 4in]{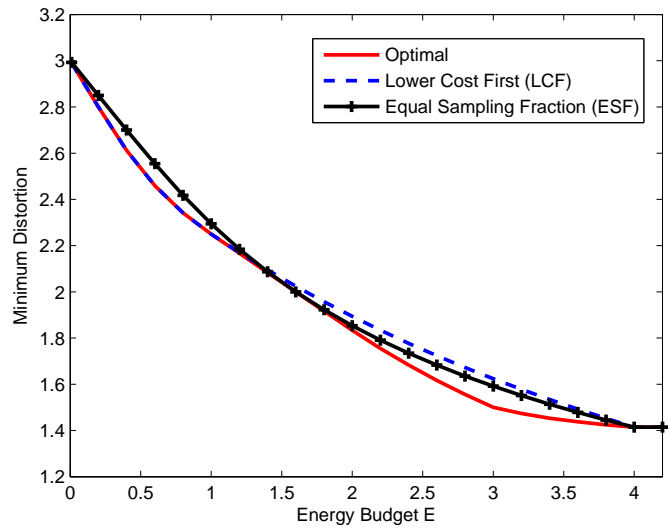}
\caption{Distortion v.s. Energy, where $R = 1$, $q_1 = q_2 = 1$, $\epsilon_{S,1} = 3$, $\epsilon_{S,2}=1$, $\sigma_1^2 = 2$ and $\sigma_2^2 = 1$.}
\label{fig:distvsE}
\end{figure}

Fig. \ref{fig:distvsE} plots the minimum distortion as a function of the energy budget $E$ when there are two classes of Gaussian sources with one source in each class, i.e., $K=2$ and $q_1 = q_2 = 1$. The parameters are chosen as $R = 1$, $\sigma_1^2 = 2$, $\sigma_2^2 = 1$, $\epsilon_{S,1} = 3$ and $\epsilon_{S,2}=1$. For the given parameters, unlike the assumptions of Proposition 1, source 1 has both larger variance and a higher sensing cost than source 2. For comparison, we consider the following two suboptimal schemes: 1) \textit{Lower Cost First} (LCF): sources are sampled starting from the one with lower cost, i.e., source 2, so that source 1 is sampled only when there is additional energy left after source 2 is fully sampled. The available rate $R$ is then split among the sensed sources using the variant of the reverse water-filling solution discussed in the previous subsection that accounts for the fact that sources may be partially sampled (recall (\ref{eqn:Rm})); 2) \textit{Equal Sampling Fraction} (ESF): the energy constraint is ignored at first and the total rate $R$ is allocated to the sources using the classic reverse water-filling solution. The available energy is then used to sample an equal fraction from the sources that have received positive rate by the reverse water-filling procedure.  Finally, with the sampling fraction of each source known, the rate $R$ is re-distributed among the sensed sources using the variant of reverse water-filling discussed in the previous subsection.

\begin{figure}
\centering
\includegraphics[width = 4in]{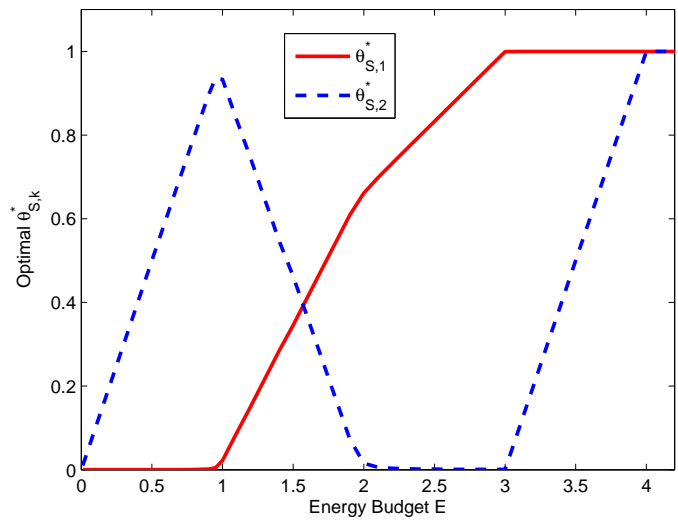}
\caption{Optimal sampling fractions $\boldsymbol\theta_S^*$ for $R = 1$, $q_1 = q_2 = 1$, $\epsilon_{S,1} = 3$, $\epsilon_{S,2}=1$, $\sigma_1^2 = 2$ and $\sigma_2^2 = 1$. }
\label{fig:thetas}
\end{figure}

Fig. \ref{fig:distvsE} shows that, when $E$ grows beyond $q_1\epsilon_{S,1}+q_2\epsilon_{S,2}= 4$, the distortion cannot be further reduced since both sources are fully sampled. Moreover, LCF is optimal for small energy budgets $E$ but becomes strictly suboptimal when $E$ grows larger than $1.1$. In this regime, ESF tends to perform better. To gain more insight into this result, the optimal $\theta_{S}^*$ is plotted in Fig. \ref{fig:thetas}. As shown in Fig. \ref{fig:thetas}, when $E$ is smaller than 0.9, all the energy is dedicated to sensing class 2. This implies that for small energy budgets $E$, sensing cost is the dominant factor in determining how energy is allocated for sensing. Instead, for $E$ larger than 1.6, a larger fraction from class 1 is sampled than from class 2, which suggests that, as $E$ increases, the variance gradually becomes a more influential factor in determining the optimal sampling fractions. This explains why ESF can outperform LCF for sufficiently large $E$.
\vspace{-0.1cm}
\section{Joint Sensing and Communication: The Ordered Variance/Cost/Noise Case}
\vspace{-0.1cm}In Section III-A, we investigated the optimal solution to the separate sensing and communication problem in (\ref{eqn:optprob}) when source variances and sensing costs are ordered. In this section, we analyze the joint sensing/communication problem in (\ref{eqn:optprob2}) when the source variances, the sensing costs and the channel noise variances are ordered. Similar to Section III-A, we divide the $Q$ parallel source-channel pairs into $K$ classes, with class $k$ having $q_k$ pairs, where $1\leq k\leq K$. It is assumed that, in class $k$, the sources have the same variance $\sigma_k^2$ and the channels have the same noise variance $N_k$. Following Section III-A, we assume the source variances and the sensing costs satisfy $\sigma_1^2 >...>\sigma_K^2$ and $\epsilon_{S,1} \leq ... \leq\epsilon_{S,K}$, respectively. It is also assumed that the channel noise variances satisfy $N_1\leq ... \leq N_K$. While for the general case, the problem in (\ref{eqn:optprob2}) can be shown to be convex, similar to the problem in (\ref{eqn:optprob}) as discussed in Section III-B, there is no closed form solution. However, for the ordered case described above, finding an analytical solution in closed form is possible under certain conditions. Similar to Section III-A, it can be readily shown that it is optimal to allocate the same sampling fraction $\theta_{S,k}$ and the same transmit power $P_k$ to all source-channel pairs in class $k$.

\begin{figure}
\centering
\includegraphics[width = 4in]{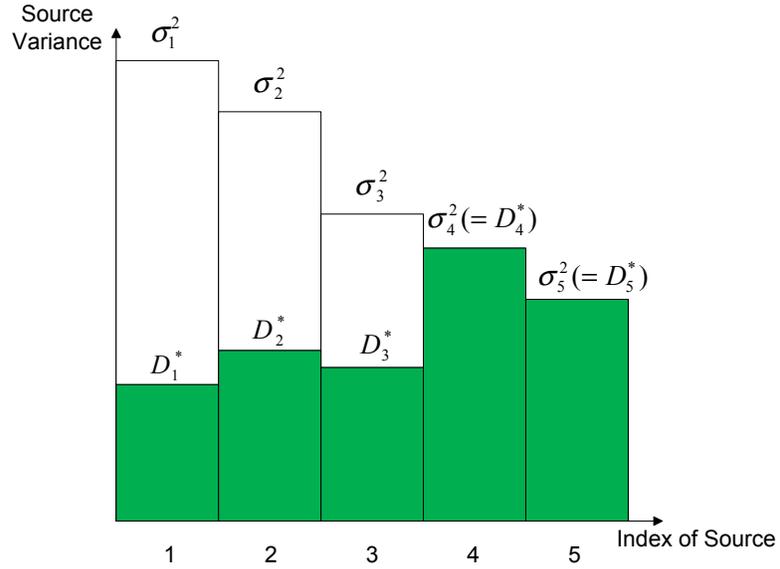}
\caption{Illustration of the optimal solution in Lemma 1 when $K=5$, $q_k = 1$, $k=1,...,5$, and $B\in \mathcal{B}_3$. For the first three sources, the optimal distortion level $D_k^*$ is proportional to $(\sigma_k^2 N_k^{2\tau})^{\frac{1}{2\tau+1}}$, for $1\leq k\leq 3$. The last two sources are assigned a zero rate and thus their distortion levels are equal to the source variances.}
\label{fig:rwf2}
\end{figure}

For convenience, we divide the range of $B$ to a sequence of intervals: $\mathcal{B}_1 = (b_0,b_1]$, $\mathcal{B}_2 = (b_1,b_2]$,..., $\mathcal{B}_K = (b_{K-1},b_K)$, where $b_0 = 0$, $b_K = +\infty$, and
\vspace{-0.2cm}
\begin{equation}
 b_i =  \tau\sum_{j=1}^{i} q_j N_j\left(\left(\frac{\sigma_j^2N_{i+1}}{\sigma_{i+1}^2N_j}\right)^{\frac{1}{2\tau+1}}-1\right) , \quad 1\leq i\leq K-1. \label{eqn:bn}\vspace{-0.2cm}
\end{equation}
We now first summarize the solution of (\ref{eqn:optprob2}) in the special case of zero sensing costs, i.e., when $\epsilon_{S,k}=0$ for all $1\leq k\leq K$. In this case, we can sample all the sources fully, i.e, we set $\theta_{S,k}=1$ for all $1\leq k\leq K$, without loss of optimality.
\vspace{-0.1cm}{\lemma \label{thm:cor1} For $K\geq 2$, assuming $\sigma_1^2 > ...> \sigma_K^2$, $N_1\leq ... \leq N_K$ and $\epsilon_{S,1}= ... =\epsilon_{S,K}=0$, if $B\in \mathcal{B}_m$ for some $1\leq m\leq K$, then it is optimal to assign positive transmit powers only to the first $m$ classes of source-channel pairs as
\vspace{-0.2cm}
\begin{equation}
P_k^* =\frac{B+ \tau\sum_{j=1,j\neq k}^m q_jN_j \left(1-\left(\frac{\sigma_j^2 N_k}{\sigma_k^2 N_j}\right)^{\frac{1}{2\tau+1}}\right)}{\tau\sum_{j=1}^m q_j \left(\frac{\sigma_j^2 N_j^{2\tau}}{\sigma_k^2 N_k^{2\tau}}\right)^{\frac{1}{2\tau+1}}}, \quad 1\leq k\leq m,
\label{eqn:P_k4}\vspace{-0.2cm}
\end{equation}
and to assign zero power to the remaining classes, i.e., $P_k^* = 0$, for $m+1\leq k\leq K$.
}
\begin{proof}
With $\theta_{S,1} = ...=\theta_{S,K} = 1$, the optimization of powers $\mathbf{P}$ in (\ref{eqn:optprob2}) is convex and can be easily performed using the standard Lagrangian approach (see also discussion below).
\end{proof}
To interpret the solution in Lemma 1, we observe, similar to Section III, the optimal power allocation can be obtained by solving the dual problem
\vspace{-0.2cm}
\begin{equation}
\min_{P_k\geq 0} \left(\sigma_k^2 \left(1+\frac{P_k}{N_k}\right)^{-2\tau} + \beta\tau P_k\right)\label{eqn:int_rwf2},\vspace{-0.2cm}
\end{equation}
for each class $k$, where $\beta$ is the Lagrangian multiplier (or ``power price'') to be selected such that $\sum_{k=1}^K q_k \tau P_k = B$ is satisfied. It can be seen that the solution to this problem is given by
\vspace{-0.2cm}
\begin{equation}
P_k^* = \left(\beta^{-\frac{1}{2\tau+1}}(2\sigma_k^2N_k^{2\tau})^{\frac{1}{2\tau+1}}-N_k\right)^+,\label{eqn:Pk}\vspace{-0.25cm}
\end{equation}
and that the corresponding achieved distortion for each of the $m$ classes that are assigned with a positive transmit power is given by
\vspace{-0.2cm}
\begin{equation}
D_k^* = \sigma_k^2\left(1+\frac{P_k^*}{N_k}\right)^{-2\tau} = \left(\frac{\beta}{2}\right)^{\frac{2\tau}{2\tau+1}}(\sigma_k^2 N_k^{2\tau})^{\frac{1}{2\tau+1}},\quad 1\leq k\leq m,\vspace{-0.25cm}\end{equation}
where $m$ is such that $B\in \mathcal{B}_m$. It is interesting to note that, in general, unlike the reverse water-filling solution, all the source-channel pairs that are allocated positive powers (or positive rates for reverse water-filling) are not assigned the same distortion level in the joint sensing and communication problem considered here. Instead, the distortion level is proportional to $(\sigma_k^2 N_k^{2\tau})^{\frac{1}{2\tau+1}}$. This shows that only in the special case of $\sigma_k^2 N_k^{2\tau}$ equal to a constant for all $1\leq k\leq m$, all the source-channel pairs with positive powers have the same distortion. Fig. \ref{fig:rwf2} illustrates an example for $K=5$ and $q_k = 1$, $1\leq k\leq 5$ with $B\in \mathcal{B}_3$.

In the case of nonzero sensing costs, it is difficult to obtain an analytical characterization even in the ordered case. Below, we first summarize some structural properties of the optimal solution and then characterize the solution when the energy budget $B$ is sufficiently large.
\vspace{-0.1cm}
{\proposition For $K\geq 2$, assuming $\sigma_1^2 > ...> \sigma_K^2$, $0<\epsilon_{S,1} \leq ... \leq\epsilon_{S,K}$ and $N_1\leq ... \leq N_K$, it is optimal to sense and transmit only the first $m$ source classes, for some $m$ with $1\leq m\leq K$ depending on the energy budget $B$. Moreover, for the sensed $m$ classes, the sampling fractions satisfy $0<\theta_{S,m}^*\leq ...\leq \theta_{S,1}^*\leq 1$, with $\theta_{S,i}^* = \theta_{S,j}^*$ ($1\leq i<j\leq m$) only when both are 1.   \label{thm:prop2}
}\vspace{-0.1cm}
\begin{proof}
The structural results on the optimal solution are obtained using the KKT conditions. As in the proof of Proposition 1, special care must be taken since the objective function in (\ref{eqn:optprob2}) is not continuously differentiable in the entire feasible set. See Appendix B for details.
\end{proof}
Proposition 2 suggests that the sources with larger variances are sampled for a fraction greater than or equal to that of the sources with smaller variances. However, unlike the separate sensing/communication scenario, the sources with larger variances do not need to be fully sampled before the sources with smaller variances are sampled.

We next characterize the optimal solution for the special case when $B$ is sufficiently large so that all sources can be fully sensed. We also compute the minimum energy budget that guarantees this. To this end, let us define the set $\bar{\mathcal{B}}$ as $\bar{\mathcal{B}} = \left[\bar{b},+\infty\right)$, where $\bar{b}$ is the solution to the equation
\begin{equation}
\frac{\sigma_K^2}{\epsilon_{S,K}}\left(1-\left(1+\frac{\bar{P}_K}{N_K}\right)^{-2\tau}\left[1+2\tau\ln\left(1+\frac{\bar{P}_K}{N_K}\right)\right]\right)=\left(\frac{\tau\sum_{j=1}^K q_j(2\sigma_j^2 N_j^{2\tau})^{\frac{1}{2\tau+1}}}{\bar{b}-\sum_{j=1}^K q_j(\epsilon_{S,j}-\tau N_j)}\right)^{2\tau+1} ,\label{eqn:prop2_ineq1}
\end{equation}
with
\begin{equation}
\bar{P}_K = \frac{\bar{b}-b_{K-1}-\sum_{j=1}^{K} q_j \epsilon_{S,j}}{\tau\sum_{j=1}^K q_j \left(\frac{\sigma_j^2 N_j^{2\tau}}{\sigma_K^2 N_K^{2\tau}}\right)^{\frac{1}{2\tau+1}}}.\label{eqn:barP}\vspace{-0.25cm}
\end{equation}
Note that with $\bar{b}\geq b_{K-1}+\sum_{j=1}^{K} q_j \epsilon_{S,j}$, the solution to (\ref{eqn:prop2_ineq1}) is unique, since over this range, the left side of (\ref{eqn:prop2_ineq1}) is a strictly increasing function of $\bar{b}$, while the right side is a strictly decreasing function of $\bar{b}$.
{\proposition For $K\geq 2$, assuming $\sigma_1^2 > ...> \sigma_K^2$, $0<\epsilon_{S,1} \leq ... \leq\epsilon_{S,K}$ and $N_1\leq ... \leq N_K$, if $B\in \bar{\mathcal{B}}$, it is optimal to fully sample all the $K$ classes of sources, i.e., to set $\theta_{S,k}^* = 1$ for all $1\leq k\leq K$ and to select transmit powers $\mathbf{P}^*$ as
\begin{equation}
P_k^* = \frac{B-\sum_{j=1}^K q_j \epsilon_{S,j}+\tau\sum_{j=1,j\neq k}^K q_jN_j \left(1-\left(\frac{\sigma_j^2 N_k}{\sigma_k^2 N_j}\right)^{\frac{1}{2\tau+1}}\right)}{\tau \sum_{j=1}^K q_j \left(\frac{\sigma_j^2 N_j^{2\tau}}{\sigma_k^2 N_k^{2\tau}}\right)^{\frac{1}{2\tau+1}}} , \quad 1\leq k\leq K. \label{eqn:Pkstar}\vspace{-0.25cm}
\end{equation}
\label{thm:prop3}}
\vspace{-0.25cm}\begin{proof}
The proof is based on that of Proposition 2 and is provided in Appendix C.
\end{proof}
Proposition \ref{thm:prop3} states that, if the energy budget is larger than the threshold $\bar{b}$, then it is optimal to fully sample all the sources and to allocate power as for the case with no sensing costs (see (\ref{eqn:P_k4})) but with energy budget discounted by the energy needed for sensing (i.e., with energy $B-\sum_{j=1}^{K} q_j \epsilon_{S,j}$). It is interesting to note that the threshold $\bar{b}$ is strictly larger than $b_{K-1}+\sum_{j=1}^{K} q_j \epsilon_{S,j}$. We recall that $b_{K-1}$ is the energy threshold above which it is optimal to assign positive powers to all $K$ classes of source-channel pairs in the zero sensing cost case, while $\sum_{j=1}^{K} q_j \epsilon_{S,j}$ is the total sensing energy needed to sense all the sources.

\begin{figure}
\centering
\includegraphics[width = 4in]{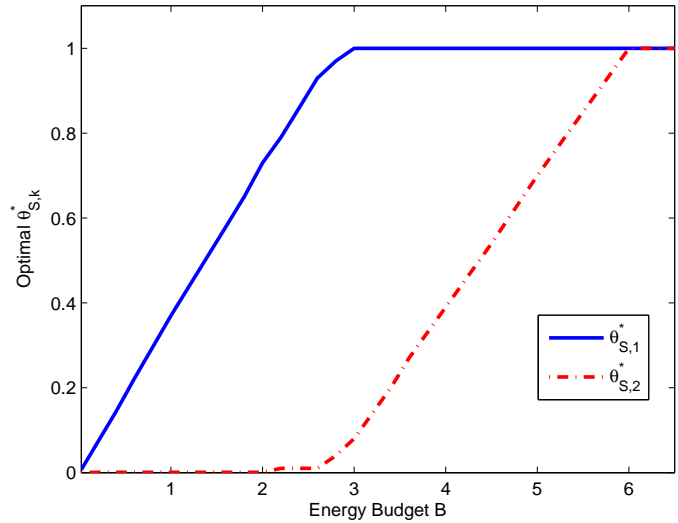}
\caption{Optimal sampling fractions $\boldsymbol\theta_{S}$ for $q_1 = q_2 = 1$, $\sigma_1^2 = 1.25$, $\sigma_2^2 = 1$, $\epsilon_{S,1} = \epsilon_{S,2}=1$ and $N_1 = N_2 = 4$.}
\label{fig:prob2_thetas1}
\end{figure}

Fig. \ref{fig:prob2_thetas1} shows the optimal sampling fractions for the joint sensing/communication problem as a function of energy budget $B$ when parameters are chosen as $q_1 = q_2 = 1$, $\sigma_1^2 = 1.25$, $\sigma_2^2 = 1$, $\epsilon_{S,1} = \epsilon_{S,2}=1$ and $N_1 = N_2 = 4$. The results are obtained via numerical methods \cite{references:Bertsekas}. It can be seen from Fig. \ref{fig:prob2_thetas1}, for any $B$, $\theta_1^*$ is greater than or equal to $\theta_2^*$, which is consistent with the optimal structure derived in Proposition \ref{thm:prop2}. Moreover, when $2<B<3$, both sources are partially sampled, which is not encountered in the optimal solution of the separate sensing and communication problem of Section III. As $B$ grows beyond $6$, both classes are fully sampled. This threshold corresponds to threshold $\bar{b}$ in (\ref{eqn:prop2_ineq1}) with $K = 2$ and is strictly larger than $b_1 + q_1\epsilon_{S,1} + q_2\epsilon_{S,2} = 2.3$. It can be observed from Fig. \ref{fig:prob2_thetas1} that, if $2.3 <B <6$, the optimal solution entails partial sampling of at least source 2 which has the lower variance. In this case, fully sampling both sources is strictly suboptimal.
\vspace{-0.25cm}
\section{Conclusions}
\vspace{-0.25cm}
In this paper, we studied an energy-constrained integrated sensor system that has a constant sensing energy cost per source sample and we investigated the impact of the sensing energy cost on the end-to-end distortion of parallel Gaussian sources. We formulated a distortion minimization problem with either separate constraints on the sensing energy budget and on the communication rates, or a joint constraint on the energy budget for both sensing and transmission. For both problems, we studied the special case in which sources with larger variances have lower sensing costs. We showed that, for the separate sensing/communication problem, the optimal strategy is to sense the sources starting from the one with the largest variance and to allocate the communication rate using reverse water-filling, or a variant of it, on the sensed sources. Moreover, for the joint sensing/communication problem, it is generally optimal to sense, possibly partially, only a subset of the sources with the largest variances and to allocate the transmit powers among their respective channels. When the source variances and the sensing costs are arbitrarily ordered, the optimal solution is obtained numerically for the first problem and compared with several suboptimal strategies. Future work includes extension of the analysis presented here to the case of an energy neutral sensor system with energy-harvesting capabilities\cite{processingcost:Ho_Zhang_11archive}. It is also of practical interest to consider more accurate models for the sensing energy cost that, for instance, account for energy costs that depend on the compression rate and the target distortion level (see, e.g., \cite{processingcost:Castiglione_Simeone_Erkip_Zemen11}).
\appendices\vspace{-0.35cm}
\section{Proof of Proposition 1}\vspace{-0.35cm}
\subsection{Overview of the Proof}\vspace{-0.35cm}
We first note that the objective function $D(\boldsymbol\theta_S,\mathbf{R})$ is convex in the set $\theta_{S,k}\geq 0$ and $R_k\geq 0$ since it is the weighted sum of convex functions $f(\theta_{S,k},R_k)$. Function $f(\theta_{S,k},R_k)$ can be easily seen to be convex since it is the linear combination of an affine function and of the perspective function of $2^{-2R_k}$ \cite{processingcost:Boyd_Vandenberghe}\footnote{Note that, in order to extend the convexity to the set $\theta_{S,k}\geq 0$ and $R_k\geq 0$, from the set $\theta_{S,k}>0$ and $R_k\geq 0$ on which convexity is guaranteed by the properties of the perspective function \cite{processingcost:Boyd_Vandenberghe}, we have used the continuity of function $f(\theta_{S,k},R_k)$ over the set $\theta_{S,k}\geq 0$ and $R_k\geq 0$ as per definition given in Section II-A.}. However, function $D(\boldsymbol\theta_S,\mathbf{R})$ is not continuously differentiable at points with $\theta_{S,k} = 0$ for any $k$.\footnote{It can be seen that, even when redefining the first-order derivative of $f(\theta_{S,k},R_k)$ at $\theta_{S,k}=0$ as being equal to the limit $\lim_{\theta_{S,k}\rightarrow 0^+} \partial f(\theta_{S,k},R_k)/\partial \theta_{S,k}$, the derivative would still be discontinuous at $R_k = 0$. In fact, we have $\lim_{\theta_{S,k}\rightarrow 0^+} \partial f(\theta_{S,k},R_k)/\partial \theta_{S,k} = 0$ for $R_k=0$ and $\lim_{\theta_{S,k}\rightarrow 0^+} \partial f(\theta_{S,k},R_k)/\partial \theta_{S,k} = -1$ for $R_k>0$.}

It is easily seen that the constraint set of function $D(\boldsymbol\theta_S,\mathbf{R})$ is a polytope. Since the function is continuous over the polytope, by Weierstrass's Theorem\cite{references:Bertsekas}, a global minimum exists. Due to convexity, locally optimal points of function $D(\boldsymbol\theta_S,\mathbf{R})$ are also globally optimal. Moreover, the constraints are affine, and thus by Slater's condition, strong duality holds and optimal Lagrange multipliers exist for the dual problem. Note that, this is true irrespective of the lack of differentiability. To find locally optimal points, we can involve the KKT conditions as being necessary and sufficient wherever the function is continuously differentiable. In particular, any point in the constraint set with $\boldsymbol\theta_S>\mathbf{0}$ (i.e., $\theta_{S,k}>0$ for all $k$) that satisfies the KKT conditions is optimal. In Appendix A-B, we show that, if $(E,R)\in A_1$, where $A_1 = \{(E,R)|E>e_{K-1},R>r_{K-1}\}$, then such a locally minimum point exists and is given by $\theta_{S,k}^* = 1$ for $1\leq k\leq K-1$, $\theta_{S,K}^* = \min((E-e_{K-1})/(q_K \epsilon_{S,K}),1)$ and $R_k^*$ as in (\ref{eqn:R_k3}) with $m$ replaced by $K$ for $1\leq k\leq K$. It is also shown that, for $E$ and $R$ such that $E>e_l$ and $R\in (r_{l-1},r_l]$ for some $1\leq l\leq K-1$, there exists points with $\boldsymbol\theta_S>\mathbf{0}$  that satisfies the KKT conditions. However, in this case, the optimal $\mathbf{R}^*$ is given by $R_k^*$ as in (\ref{eqn:R_k4})  for $1\leq k\leq l$, and $R_k^* = 0$ for $l+1\leq k\leq K$, therefore, as long as $\theta_{S,k}^* = 1$ for all $1\leq k\leq l$ and $\sum_{k=l+1}^K q_k\theta_{S,k}^*\leq E-e_l$, the choice of $\theta_{S,k}^*$ for $l+1 \leq k\leq K$ is arbitrary. For all other choices of $(E,R)$, no point with $\boldsymbol\theta_S>\mathbf{0}$ satisfies the KKT conditions and thus the optimal solution must have some sampling fractions $\theta_{S,k}^*$ equal to zero. It is not hard to see that, due to the order imposed on the variances and sensing costs, in such a case, $\theta_{S,K}^*$ must be set to zero. Hence, we can conclude that if $(E,R)\notin A_1$, then it is optimal to set $\theta_{S,K}^* = 0$ and $R_K^* = 0$. The problem then reduces to the one studied above but with only the first $K-1$ classes of sources. Therefore, an optimal solution of this problem can be obtained by again solving the KKT conditions. By using the same reasoning as above, an optimal solution is found only if $(E,R)$ belongs to $A_2 = \{(E,R)|(E,R)\notin A_1, E>e_{K-2}, \text{ and } R>r_{K-2}\}$ or $E>e_l$ and $R\in (r_{l-1},r_l]$ for some $1\leq l\leq K-2$. If such conditions are not met, then the optimal solution must have $\theta_{S,K}^*=\theta_{K-1}^*=0$ and $R_K^* = R_{K-1}^* = 0$. The procedure is repeated until a solution is found by solving the KKT conditions. Note that, as mentioned, the optimal solution must exist by Weierstrass's theorem.
\vspace{-0.25cm}
\subsection{Solving the KKT Conditions}
\vspace{-0.3cm}
To find whether an optimal point exists with $\boldsymbol\theta_{S}>\mathbf{0}$, we define the Lagrangian function
\vspace{-0.2cm}
\begin{align}
L_1(\boldsymbol\theta_S, \mathbf{R},\boldsymbol\mu,\boldsymbol\nu,\alpha,\beta) &= \sum_{k=1}^K \sigma_k^2 q_k f(\theta_{S,k},R_k) + \sum_{k=1}^K \mu_k (\theta_{S,k}-1)\\
& + \sum_{k=1}^K \nu_k (-R_k) + \alpha \left(\sum_{k=1}^K q_kR_k - R\right) + \beta \left(\sum_{k=1}^K\theta_{S,k}q_k\epsilon_{S,k}-E\right),\vspace{-0.3cm}
\end{align}
and invoke the KKT conditions which are both necessary and sufficient \cite{references:Bertsekas}. It follows that $(\boldsymbol\theta_{S}^*,\mathbf{R}^*)$ is an optimal point with $\boldsymbol\theta_S>\mathbf{0}$, if and only if there exists Lagrange multiplier vectors $\boldsymbol\mu^*\geq \mathbf{0}$, $\boldsymbol\nu^*\geq \mathbf{0}$ and multipliers $\alpha^*\geq 0$, $\beta^*\geq 0$ such that
\vspace{-0.2cm}
\begin{subequations}
 \begin{align}
&\frac{\partial L_1}{\partial \theta_{S,k}} = \sigma_k^2 q_k \left(-1+2^{-\frac{2R_k^*}{\theta_{S,k}^*}}\left(1+\frac{(2\ln2)R_k^*}{\theta_{S,k}^*}\right)\right) + \mu_k^* + \beta^* q_k\epsilon_{S,k}^* = 0, \quad k = 1,2,...,K,\label{eqn:derivative0} \\\vspace{-0.1cm}
&\frac{\partial L_1}{\partial R_k} = -(2\ln2)\sigma_k^2 q_k 2^{-\frac{2R_k^*}{\theta_{S,k}^*}} -\nu_k^* +\alpha^* q_k = 0, \quad k = 1,2,...,K, \label{eqn:derivative1}
\end{align}\vspace{-1.5cm}
\end{subequations}
\begin{subequations}
 \begin{align}
  &\mu_k^*(\theta_{S,k}^*-1)=0,\; \nu_k^*(-R_k^*)=0, \quad k = 1,...,K, \label{eqn:muk_nuk1}\\\vspace{-0.1cm}
  &\alpha^*\left(\sum_{k=1}^K q_k R_k^*-R\right)=0,   \label{eqn:cons_sum_R}\\\vspace{-0.1cm}
  \text{and}\quad &\beta^*\left(\sum_{k=1}^K \theta_{S,k}^*q_k \epsilon_{S,k}-E\right)=0\label{eqn:cons_E}\vspace{-0.3cm}
\end{align}\label{eqn:kkt1}\end{subequations}are satisfied. It can be seen that we can find a solution only in the following cases.
\begin{itemize}
\item Case 1: $\mathbf{R}^*$ satisfies $R_k^*>0$ for $1\leq k\leq K$ while $\boldsymbol\theta_S^*$ satisfies $\theta_{S,k}^* = 1$ for $1\leq k\leq K-1$ and $0<\theta_{S,K}^*<1$. It is easily seen that for these to hold, we need $E>e_{K-1}$. By (\ref{eqn:muk_nuk1}), $0<\theta_{S,K}^*<1$ implies $\mu_K^* = 0$ and thus it follows from (\ref{eqn:derivative0}) that $\beta^*>0$. Then, by (\ref{eqn:cons_E}), $\sum_{k=1}^K \theta_{S,k}^*q_k \epsilon_{S,k}=E$ holds. Therefore, for $E< \sum_{i=1}^K q_i \epsilon_{S,i}$, we have $\theta_{S,K}^* = (E-e_{K-1})/(q_K\epsilon_{S,K})$. From (\ref{eqn:muk_nuk1}), $\nu_k^* = 0$ holds for any $k$. Also, it follows from (\ref{eqn:derivative1}) that $\alpha^* >0$ and $R_k^* = (\theta_{S,k}^*/2)\log_2((2\ln 2)\sigma_k^2/\alpha^*)$ for 1$\leq k\leq K$. By (\ref{eqn:cons_sum_R}), $\alpha^*>0$ implies $\sum_{k=1}^K q_k R_k^*=R$. Thus, we obtain $\alpha^* = (2\ln2)2^{-2(R - \frac{1}{2}\sum_{j=1}^K q_j\theta_{S,j}^* \log_2 \sigma_j^2)/\sum_{j=1}^K q_j\theta_{S,j}^*}$
 and $\mathbf{R}^*$ as in (\ref{eqn:R_k3}) with $m$ replaced by $K$. It is easily seen that in order to have $R_K^*>0$ we need $R>r_{K-1}$. Hence, there exists a valid solution in this case if and only if $e_{K-1}<E<\sum_{i=1}^K q_i \epsilon_{S,i}$ and $R>r_{K-1}$.
\item Case 2: For some $1\leq l\leq K$, $\mathbf{R}^*$ satisfies $R_k^*>0$ for $1\leq k\leq l$ and $R_k^* = 0$ for $l+1\leq k\leq K$, while $\boldsymbol\theta_S^*$ satisfies $\theta_{S,k}^* = 1$ for $1\leq k\leq l$ and $0<\theta_{S,k}^*<1$ for $l+1\leq k\leq K$. For these conditions to hold, $E$ needs to satisfy $E>e_l$ if $1\leq l\leq K-1$ or $E\geq \sum_{i=1}^K q_i\epsilon_{S,i}$ if $l = K$. Similar to Case 1, we obtain $\alpha^* = (2\ln2)2^{-2(R - \frac{1}{2}\sum_{j=1}^l q_j\log_2 \sigma_j^2)/(\sum_{j=1}^l q_j)}$ and $R_k^*$ as in (\ref{eqn:R_k4}) for $1\leq k \leq l$. If $l\leq K-1$, $\nu_{l+1}^*\geq 0$ implies $R\leq r_l$. If $l\geq 2$, $R_l^*>0$ implies $R>r_{l-1}$. Hence, there exists a valid solution if and only if $E>e_l$ and $R\in (r_{l-1}, r_l]$ for $1\leq l\leq K-1$ or $E\geq \sum_{i=1}^K q_i \epsilon_{S,i}$ and $R>r_{K-1}$.
\end{itemize}
We observe from the above analysis that only when $(E,R)$ belongs to $A_1=\{(E,R)|E>e_{K-1}, R>r_{K-1}\}$, there exists a unique optimal solution to the KKT conditions with $\boldsymbol\theta_{S}>\mathbf{0}$, which is given by $\theta_{S,k}^* = 1$ for $1\leq k\leq K-1$, $\theta_{S,K}^* = \min((E-e_{K-1})/(q_K\epsilon_{S,K}),1)$ and $\mathbf{R}^*$ as in (\ref{eqn:R_k3}) with $m$ replaced by $K$. For $E$ and $R$ such that $E> e_l$ and $R\in (r_{l-1},r_l]$ for some $1\leq l\leq K-1$, there also exists points with $\boldsymbol\theta_{S}>\mathbf{0}$ that satisfy the KKT conditions and at these points the optimal rate allocation $\mathbf{R}^*$ is given by $R_k^*$ as in (\ref{eqn:R_k4}) for $1\leq k\leq l$, and $R_k^* = 0$ for $l+1\leq k\leq K$. Following the discussion in Appendix A-A, this concludes the proof.
\vspace{-0.5cm}
\section{Proof of Proposition \ref{thm:prop2}}
\vspace{-0.5cm}
\subsection{Overview of the Proof}
\vspace{-0.25cm}
Similar to function $D(\boldsymbol\theta_S,\mathbf{R})$ in Appendix A, it can be shown that function $D(\boldsymbol\theta_S, \mathbf{P})$ is convex but not differentiable at points with $\theta_{S,k} = 0$ for any $k$. Moreover, to obtain optimal points, we can invoke the KKT conditions as being necessary and sufficient wherever function $D(\boldsymbol\theta_S,\mathbf{P})$ is continuously differentiable. Therefore, as in Appendix A, any point in the constraint set with $\boldsymbol\theta_{S}>\mathbf{0}$ that satisfies the KKT conditions is optimal. It is shown in Appendix B-B that, if a minimum point with $\boldsymbol\theta_{S}>\mathbf{0}$ exists, it has the structure that $0<\theta_{S,K}^*\leq ...\leq \theta_{S,1}^*\leq 1$ and $P_k^*>0$ for $1\leq k\leq K$, with $\theta_{S,i}^*=\theta_{S,j}^*$ ($1\leq i<j\leq K$) only when both are equal to 1. Instead, if no point with $\boldsymbol\theta_S>\mathbf{0}$ satisfies the KKT conditions, similar to Appendix A, we must have $\theta_{S,K}^* = 0$ and accordingly $P_K^* = 0$. The problem is then effectively reduced to the one studied above but with only the first $K-1$ classes of source-channel pairs . Using the same reasoning as above, if a minimum point with $\theta_{S,k}^*>0$ for all $1\leq k\leq K-1$ exists, it must satisfy $0<\theta_{S,K-1}^*\leq ...\leq \theta_{S,1}^*\leq 1$ and $P_k^*>0$ for $1\leq k\leq K-1$; otherwise, we have $\theta_{S,K-1}^* = \theta_{S,K}^* = 0$ and $P_{K-1}^* = P_K^* = 0$. By repeating this procedure, we can find the structure of any possible optimal solution as stated in Proposition \ref{thm:prop2}.
\vspace{-0.25cm}
\subsection{Solving the KKT Conditions}
\vspace{-0.25cm}
Similar to Appendix A-B, we can define a Lagrangian function $L_2(\mathbf{\theta}_S, \mathbf{P},\boldsymbol\mu,\boldsymbol\nu,\beta)$, with $\beta$ being the Lagrangian multiplier corresponding to the total energy constraint. From the KKT conditions, it follows that $(\boldsymbol\theta_{S}^*,\mathbf{P}^*)$ is an optimal point with $\boldsymbol\theta_S>\mathbf{0}$ if and only if there exists Lagrange multiplier vectors $\boldsymbol\mu^*\geq \mathbf{0}$, $\boldsymbol\nu^*\geq \mathbf{0}$ and multiplier $\beta^*\geq 0$ such that
\vspace{-0.2cm}
\begin{subequations}
\begin{align}
&\frac{\partial L_2}{\partial \theta_{S,k}} = \sigma_k^2 q_k \left(-1+ \left(1+\frac{P_k^*}{N_k}\right)^{-\frac{2\tau}{\theta_{S,k}^*}} \left[1+\frac{2\tau}{\theta_{S,k}^*}\ln\left(1+\frac{P_k^*}{N_k}\right)\right]\right) + \mu_k^* + \beta^* q_k\epsilon_{S,k} = 0, \label{eqn:derivative0_n}\\
&\frac{\partial L_2}{\partial P_k} = \sigma_k^2 q_k \left(-\frac{2\tau}{N_k}\left(1+\frac{P_k^*}{N_k}\right)^{-\frac{2\tau}{\theta_{S,k}^*}-1}\right)  -\nu_k^* +\beta^*\tau q_k = 0, \quad k = 1,2,...,K, \label{eqn:derivative1_n}
\end{align}\label{eqn:derivative2}\vspace{-1.5cm}
\end{subequations}
\begin{subequations}
\begin{align}
&\mu_k^*(\theta_{S,k}^*-1)=0,\;\nu_k^*(-P_k^*)=0, \quad k = 1,...,K, \label{eqn:muk_nuk2}\\
\text{and}\quad &\beta^*\left(\sum_{k=1}^K q_k(\theta_{S,k}^* \epsilon_{S,k}+P_k^*)-B\right)=0\label{eqn:cons_B}\vspace{-0.2cm}
\end{align}\label{eqn:kkt2}\end{subequations}are satisfied. Given the joint energy constraint, if any optimal point with $\boldsymbol\theta_S>\mathbf{0}$ exists, it is easily seen that it must satisfy $P_k>0$ for $1\leq k\leq K$.

We now show that any $(\boldsymbol\theta_S,\mathbf{P})$, with $0<\theta_{S,i}\leq \theta_{S,j} <1$ for some $i$, $j$ satisfying $1\leq i< j\leq K$, can be ruled out as an optimal solution. In this case, $0<\theta_{S,i}\leq \theta_{S,j} <1$ yields $\mu_i = \mu_j = 0$. Moreover, from (\ref{eqn:derivative0_n}) and $\sigma_i^2/\epsilon_{S,i}> \sigma_j^2/\epsilon_{S,j}$, it follows that $(1+P_i/N_i)^{2\tau/\theta_{S,i}}<(1+P_j/N_j)^{2\tau/\theta_{S,j}}$, or $\ln(1+P_i/N_i)/\ln(1+P_j/N_j)<\theta_{S,i}/\theta_{S,j}$. Also, by (\ref{eqn:muk_nuk2}), $P_i>0$ and $P_j>0$ imply $\nu_i = \nu_j = 0$. From (\ref{eqn:derivative1_n}) and $\sigma_i^2/N_i>\sigma_j^2/N_j$, it follows that $(1+P_i/N_i)^{2\tau/\theta_{S,i}+1}>(1+P_j/N_j)^{2\tau/\theta_{S,j}+1}$, or $\ln(1+P_i/N_i)/\ln(1+P_j/N_j)>(1+2\tau/\theta_{S,j})/(1+2\tau/\theta_{S,i})$. If $\theta_{S,i}\leq \theta_{S,j}$, then $(1+2\tau/\theta_{S,j})/(1+2\tau/\theta_{S,i})\geq \theta_{S,i}/\theta_{S,j}$. Hence, we have a contradiction. Similarly, the case of $(\boldsymbol\theta_S,\mathbf{P})$ with $0<\theta_{S,i}<1$, $\theta_{S,j}=1$ for some $i$, $j$ satisfying $1\leq i< j\leq K$, can also be ruled out. Hence, any optimal point with $\boldsymbol\theta_S>\mathbf{0}$ has the following structure: $\mathbf{P}^*$ satisfies $P_k^*>0$ for all $1\leq k\leq K$, while $\boldsymbol\theta_S^*$ satisfies $0<\theta_{S,K}^*\leq...\theta_{S,2}^*\leq \theta_{S,1}^*\leq 1$ with $\theta_{S,i}^* = \theta_{S,j}^*$ for some $i\neq j$ only when both are 1. Following the discussion in Appendix B-A, this concludes the proof.
\vspace{-0.25cm}
\section{Proof of Proposition 3}
\vspace{-0.25cm}
Using the KKT conditions in (\ref{eqn:derivative2})-(\ref{eqn:kkt2}), we can derive a closed-form solution for the special case when the optimal solution satisfies $\theta_{S,k}^* = 1$ for all $1\leq k\leq K$. By Proposition \ref{thm:prop2}, it follows that in any such solution, $P_k^*>0$ for $1\leq k\leq K$. By (\ref{eqn:muk_nuk2}), $\nu_k^* = 0$ holds for any $k$. It follows from (\ref{eqn:derivative1_n}) that $\beta^*>0$ and $P_k^* =  \beta^{-\frac{1}{2\tau+1}}(2\sigma_k^2N_k^{2\tau})^{\frac{1}{2\tau+1}} - N_k$ for $1\leq k\leq K$. Also, (\ref{eqn:cons_B}) yields $\sum _{k=1}^K q_k (\epsilon_{S,k}+\tau P_k^*) = B$. Therefore, we get
\vspace{-0.2cm}
 \begin{equation}
 \beta^* = \left(\frac{\tau\sum_{k=1}^K q_k(2\sigma_k^2 N_k^{2\tau})^{\frac{1}{2\tau+1}}}{B-\sum_{k=1}^Kq_k(\epsilon_{S,k}-\tau N_k)}\right)^{2\tau+1},  \label{eqn:beta2}
 \end{equation}\vspace{-0.2cm}and $P_k^*$ as in (\ref{eqn:Pkstar}) for $1\leq k\leq K$. Note that $\mu_k^* \geq 0$ needs to hold for any $k$. With parameters ordered, it can be seen that for these conditions to hold, it is sufficient to have $\mu_K^*\geq 0$, i.e.,
 \vspace{-0.2cm}
\begin{equation}
\beta^* \leq \frac{\sigma_K^2}{\epsilon_{S,K}}\left(1-\left(1+\frac{P_K^*}{N_K}\right)^{-2\tau}\left[1+2\tau\ln\left(1+\frac{P_K^*}{N_K}\right)\right]\right).\label{eqn:beta_ineq}
\end{equation}\vspace{-0.2cm}Hence, this solution is valid if and only if $B\geq \bar{b}$ where $\bar{b}$ is as defined in Section IV and is the value of $B$ when (\ref{eqn:beta_ineq}) is met with equality. This concludes the proof.
\vspace{-0.5cm}
\bibliographystyle{IEEEtran}
\bibliography{IEEEabrv,processingcost}
\newpage
\end{document}